\pdfoutput=1

\documentclass[twocolumn,english]{revtex4}
\usepackage{times,amssymb,amsmath,graphicx}
\usepackage[bookmarks=false,pdffitwindow=false,pdfstartview={FitH}]{hyperref}
\usepackage[T1]{fontenc}
\usepackage{babel,multirow}
\usepackage{color}

\begin{document}


\title{
  Josephson network as a~model for high-temperature superconductor
  and beyond: Macroscopic quantum coherence probed by microwave absorption 
}

\author{Adam Rycerz\footnote{Correspondence: 
  \href{mailto:rycerz@th.if.uj.edu.pl}{rycerz@th.if.uj.edu.pl}.}}
\affiliation{Institute for Theoretical Physics,
  Jagiellonian University, \L{}ojasiewicza 11, PL--30348 Krak\'{o}w, Poland}

\date{September 4, 2026}

\begin{abstract}
  Since the seminal experiment by Stankowski {\em et al.\/}
  [\href{https://doi.org/10.1103/PhysRevB.36.7126}{%
  Phys.\ Rev.\ B {\bf 36} 7126 (1987)}] on YBa$_2$Cu$_3$O$_{7-x}$,
  Magnetically-Modulated Microwave Absorption (MMMA) has become an important
  technique for detecting the superconducting transition in inhomogeneous
  ceramic compounds. 
  The rise of microwave absorption below $T_c$ is accompanied by several
  low-field anomalies, usually attributed to the weak links between
  superconducting grains showing the Josephson effect, or to the dynamics
  of Abrikosov-Josephson vortices. 
  In this article, we briefly review selected theoretical models rationalizing
  MMMA spectra for high-temperature superconductors, focusing on the
  $3$-dimensional array of Josephson junctions with random parameters
  including the resistivity, capacity and inductance of each junction,
  showing characteristic absorption anomalies observed in the experiment.
  The implications for recently studied microwave absorption in Josephson
  junction qubits are also discussed. 
\end{abstract}

\maketitle


\section{Introduction}
After the discovery of nitrogen-temperature superconductivity in
copper oxides \cite{Bed86,MWu87} it became  clear that standard signatures
of the superconducting phase (i.e., zero resistivity accompanied by the
Meissner-Ochsenfeld effect) are difficult to identify due to the
inhomogeneous (granular) structure of these materials, in which
superconducting grains are surrounded by a~normal-state material, with weak
junctions connecting the grains \cite{Sim94}.
In the series of pioneering 1987 experiments \cite{Sta87,Dur87,Bla87},
samples of the superconductor YBa$_2$Cu$_3$O$_{7-x}$, with $x\sim{}0.1$ and the
transition temperature $T_c\approx{}91\div{}95\,$K, were placed in a~standard
EPR spectrometer, which provides the magnetic resonance response due to
Cu$^{2+}$ in the normal phase in the field range of $0.35\div{}0.40\,$T.
Below $T_c$, the lower-field signal has a~reversed phase with respect to
the usual (Lorentzian) EPR line, indicating a~large
nonresonant microwave absorption (in zero and low applied magnetic field)
that can be associated with the transition to the superconducting phase.

\begin{figure}[!b]
\includegraphics[width=\linewidth]{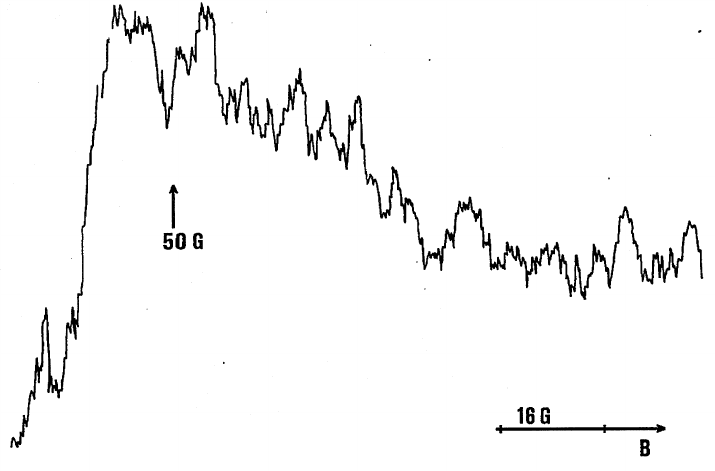}
\caption{ \label{stankowsfig}
  First derivative of the power MMMA spectrum showing the $50$-G
  peak for a~sample of the ceramic superconductor
  YBa$_2$Cu$_3$O$_{7-x}$ at $T=80\,$K.
  Reprinted from Ref.\ \cite{Sta87}. 
}
\end{figure}

An exemplary field dependence, illustrating the above-mentioned 
{\em magnetically-modulated microwave absorption} (MMMA), 
is displayed in Fig.~\ref{stankowsfig} \cite{Sta87}.
Remarkably, a~broad absorption structure is modulated by a~noisy signal with
a~clear quasiperiodic substructure.
Later, a~similar feature was identified in physical models built from
type-I superconductor grains connected by weak junctions
\cite{Dru90,Try90,Rub91}, yet did not appear in single-crystal measurements
for type-II superconductors \cite{Rub92,Bys92}. 
What is more, nonlinear phenomena, e.g., a~generation of odd harmonics in
zero applied field and of even harmonics in nonzero field were also
observed \cite{Jef88,Xia89}.

Different theoretical considerations attributed the key features of low-field
MMMA spectra for ceramic superconductors to the flux quantization in
elementary (i.e., minimal in size) loop formed by superconducting grains
\cite{Sta87}, to the power loss in the dissipative weak junctions
connecting the grains \cite{Czy91,Czy95,DeL99,MLi03} forming Josephson
networks \cite{Wol93,Rei95,Nie97,Prz97,Ryc03,Ryc07}, 
or to the vortex dynamics in the presence of perpendicular (quasi)static
and microwave fields
\cite{Cle91,Cof91,YLi91,Shi91,Bra95,Ven98,Bar04,Sha08,Cso18,Liu24}. 
Very recently, an analogy to non-linear microwave absorption in Josephson
junction qubits \cite{Kra19} has been put forward \cite{Bur26}. 

In the remaining parts of the paper, we first --- in Sec.\ \ref{batheor}
--- briefly overview selected theoretical models rationalizing the MMMA spectrum
anomalies. 
Then, in Sec.\ \ref{jos3dmod}, we present the numerical approach to 
a~general three-dimensional array of the Josephson junctions shunted by
resistance ($R$), capacitance ($C$), and with inductance ($L$), together with
inclusion of some random variations of the junction parameters.
The resulting applied magnetic field dependence of the microwave absorption
by cubic arrays of up to $30\times{}30\times{}30$ Josephson junctions are
discussed in Sec.\ \ref{numres}.
Possible implications for networks of superconducting qubits are addressed
in Sec.\ \ref{qbitnet}.
The conclusions are given in Sec.\ \ref{conclu}.

\section{Theoretical interpretations of the MMMA spectrum anomalies}
\label{batheor}

\subsection{Flux quantization in elementary loop}
Stankowski {et al.\/} \cite{Sta87} proposed a~preliminary interpretation
of quasiperiodic substructure visible in the low-field MMMA spectrum.
As superconducting grains form closed loops containing Josephson junctions
in series, see Fig.\ \ref{sketch3modsfig}(a), minimal power absorption
is expected when static magnetic field is adjusted such that a~typical
elementary loop ($\alpha$), with its crossection  $S_\alpha$, is pierced by
a~magnetic flux being an integer multiplicity of the magnetic flux quantum,
\begin{equation}
B_{\alpha}^{({\rm ext})}S_{\alpha}=n\frac{h}{2e},\ \ \ \ \ \ 
n=0,\pm{}1,\pm{}2,\dots,
\end{equation}
with $B_\alpha=\hat{\bf s}_\alpha\cdot{}{\bf B}^{({\rm ext})}$ the field component
normal to the loop (defined via the unit normal vector $\hat{\bf s}_\alpha$)
and the flux quantum $\Phi_0=h/2e\approx{}2.0678\cdot{}10^{-15}\,\text{Wb}$
in the SI units. 
In such a~case, the screening current in the loop is close to zero, the phase
difference at each junction is close to an integer
multiplicity of $\pi$, and relatively high microwave field amplitude can
be applied without exceeding the critical current.
Taking the position of the first anomaly at $B_{\alpha}^{({\rm ext})}=5\,\text{mT}$,
we obtain the typical loop size (coinciding with the grain size) 
$\sqrt{S_\alpha}\approx{}0.74\,\mu\text{m}$, being a~reasonable estimate
for YBa$_2$Cu$_3$O$_{7-x}$.

\subsection{Effective dynamics of a~single junction}
An alternative reasoning was later presented by Czy\.zak and Stankowski
\cite{Czy91} (later elaborated by Czy\.zak for a~system of few junctions in
series \cite{Czy95}), who pointed out that one can focus on a~single junction
between the two superconducting grains, for which a~typical fluctuation
of the total bias current --- during the MMMA measurement --- is of the
order of squared critical current, i.e., $\langle{}I_\mu^2\rangle\sim{}I_c^2$,
and 
the magnetic flux ($\Phi$) penetrating the junction reduces its critical
current according to
\begin{equation}
\label{icphi}
I_c(\Phi)=I_c{(0)}\left|\frac{\sin\pi\Phi/\Phi_0}{\pi\Phi/\Phi_0}\right|. 
\end{equation} 
In the framework of the {\em resistively and capacitively shunted junction
model\/} (RCSJ), see Fig.\ \ref{sketch3modsfig}(b),
the phase difference ($\theta$) follows the equation of motion
\begin{equation}
\label{eqmo1jj}
C\frac{\Phi_0}{2\pi}\ddot{\theta}+\frac{1}{R}\frac{\Phi_0}{2\pi}\dot{\theta}
+I_{c}(\Phi)\sin\theta=I_\mu(t), 
\end{equation}
where $C$ and $R$ are the junction capacity and resistance, respectively,
$\dot{\theta}$ ($\ddot{\theta}$) is the first (second) time derivative
of $\theta$, and $I_c=I_c(\Phi)$ is given by Eq.\ (\ref{icphi}).  
For low-frequency bias current $I_\mu(t)$, the role of displacement current
can be neglected at a~first approximation, so 
the power absorbed by the resistance can be approximated as
\begin{align}
  P_{\rm abs}(\Phi) &\approx{} R\langle{}I_\mu^2\rangle-R{}I_c^2(\Phi)
  \nonumber \\
  &= P_\mu'+P_0\left[
    1-\left(\frac{\sin\pi\Phi/\Phi_0}{\pi\Phi/\Phi_0}\right)^2
  \right], \label{pabsczy}
\end{align}
where we have additionally defined $P_0=R{}I_c^2(0)$ and
$P_\mu'=R_N{}\left[\langle{}I_\mu^2\rangle-I_c^2(0)\right]$.

The curve given by Eq.\ (\ref{pabsczy}) is depicted schematically in
Fig.\ \ref{sketch3modsfig}(b).
A~somewhat more cumbersome analysis, including up to $12$ junctions in series,
is capable of generating absorption spectra that are closer to the
experimental ones.

\subsection{Vortex dynamics}
\begin{figure*}[!t]
\includegraphics[width=0.7\linewidth]{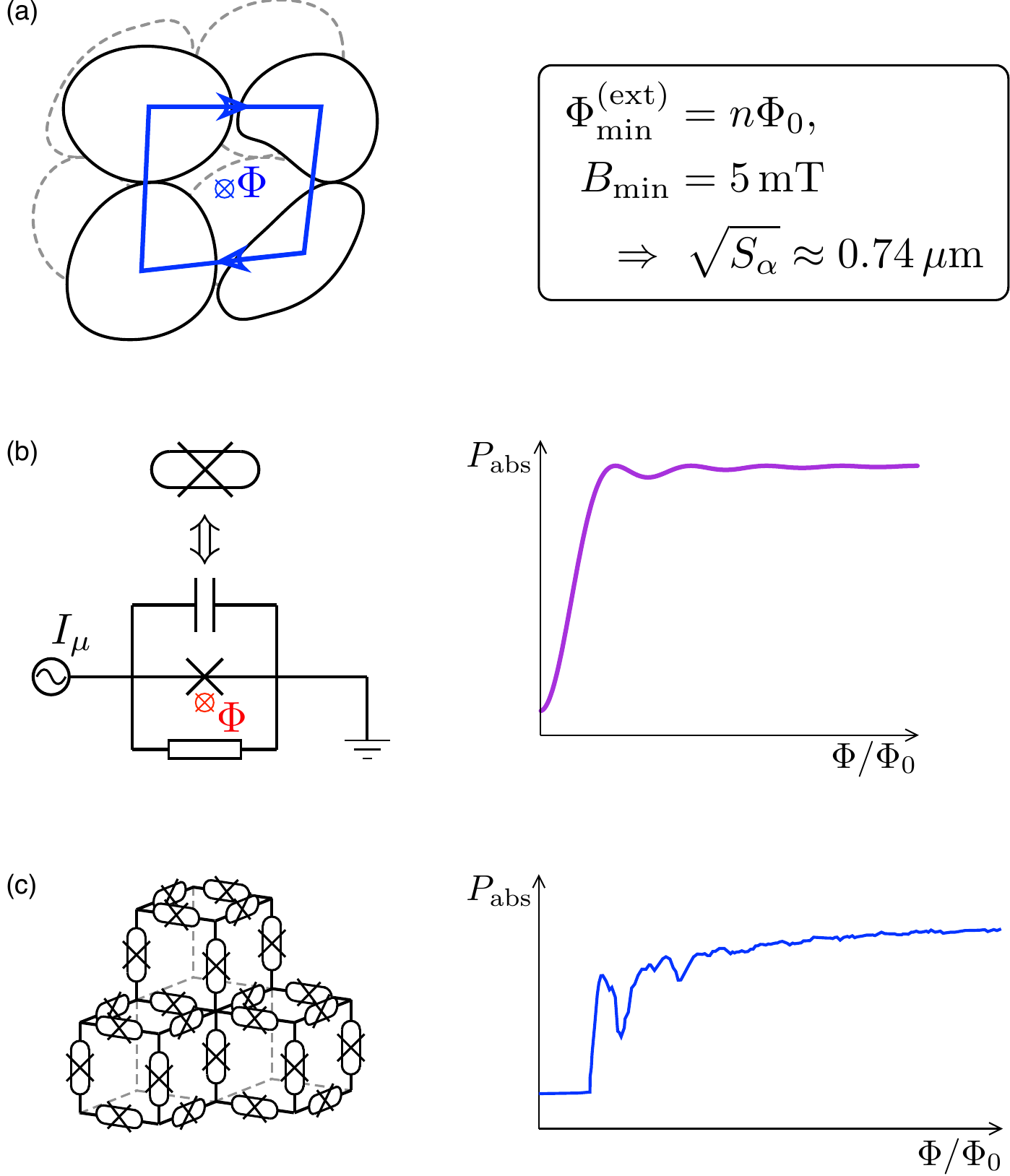}
\caption{ \label{sketch3modsfig}
Selected theoretical models allowing to understand the absorption spectrum
of Fig.\ \ref{stankowsfig}.
(a) A~superconducting loop, pierced by a~static flux $\Phi$, is expected
to show the minimal absorption for an {\em ac} signal when 
$\Phi\approx{}\Phi_{\rm min}^{({\rm ext})}=n\Phi_0$.
(b) A~single junction pierced by the flux, reducing its critical current
$I_c=I_c(\Phi)$, and subjected to the {\em ac} bias current $I_\mu$, shows
the absorbed power ($P_{\rm abs}$) that rises with $\Phi$, reaching the maximum
at $\Phi=\Phi_0$, and then showing small quasiperiodic oscillations.
(c) Three-dimensional Josephson network (schematic), containing
$N\times{}N\times{}N$ resistively and capacitively shunted junctions
with some random variation of the parameters, 
is sufficient to account for the essential experimental observations
starting from $N=30$  (see the main text for details). 
}
\end{figure*}

Preliminary discussions focusing on the loop containing Josephson
junctions as a~key building block for the model of ceramic superconductor
\cite{Sta87,Czy91,Czy95,DeL99,MLi03} has spurred numerous further
analyses of more complex Josephson networks \cite{Wol93,Rei95,Nie97,Prz97},
including the three-dimensional arrays \cite{Ryc03,Ryc07}, in attempt to
interpret the MMMA behavior (sometimes called the
{\em paramagnetic Meissner effect\/} \cite{MLi03}.
The details of the model studied in Ref.\ \cite{Ryc07} and presented
schematically in Fig.\ \ref{sketch3modsfig}(c) are revisited 
later in this paper (see Sec.\ \ref{jos3dmod}); but now, we briefly
overview main results following from the complementary approach concentrated
on the dynamics of vortices (each carrying a quantum of magnetic
flux, $\Phi_0$) in the presence of external electromagnetic field.

In 1991, Coffey and Clem \cite{Cof91} improved the Bardeen-Stephen 
model describing the viscous motion of vortices in the mixed state occurring
in high-temperature superconductors \cite{Bar65},
by including the effect of pinning force on the vortex motion.
Several related works investigated the frequency-dependent conductivity 
and dynamics of Abrikosov-Josephson vortices in the presence of both static
and {\em ac} magnetic fields \cite{YLi91,Shi91,Bra95,Ven98}.
Similar study for a~single, long Josephson junction at few-Kelvin temperatures
showed that --- in the absence of static magnetic field --- an {\em ac} bias
of a~small frequency and sufficiently large amplitude may change the
current-voltage characteristics in a~way similar to the effect of static
magnetic field in standard EPR measurements \cite{Bar04}.
The dynamics of Abrikosov-Josephson vortices in BiSCCO superconductors
with different degrees of anisotropy were also studied \cite{Sha08}, 
exhibiting a~generic magnetic field dependence of the microwave absorption,
increasing at small fields, reaching a maximum, and decreasing at large
fields. 

More recently, enhanced microwave absorption (i.e., larger than that in the
normal state) was detected in grains of type-II superconductors MgB$_2$
and K$_3$C$_{60}$ smaller than the magnetic  penetration depth \cite{Cso18},
and the interpretation in terms of the Coffey-Clem model was provided.
A~newer theoretical study showed that an additional (next to the vortex
motion) absorption mechanism, related to the quasiparticle excitations for
each vortex line, may also play a~role in microwave absorption phenomena
in type-II superconductors \cite{Liu24}.

\section{Dissipative lattice of the Josephson junctions}
\label{jos3dmod}
\begin{figure}[!t]
\includegraphics[width=0.9\linewidth]{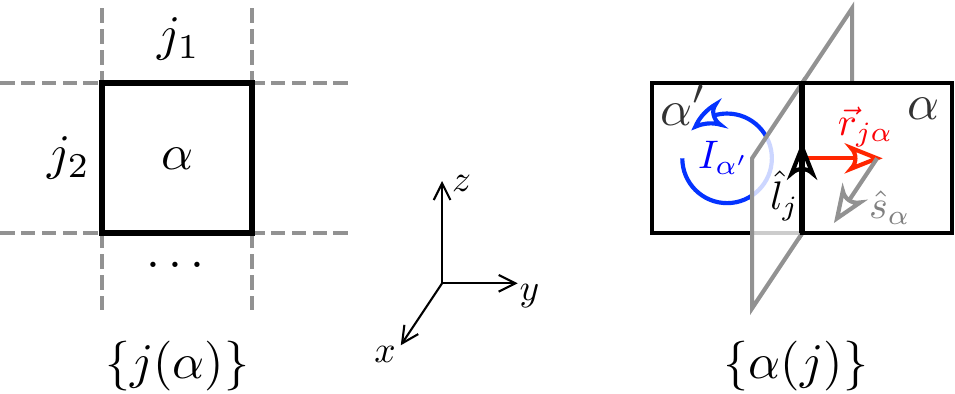}
\caption{ \label{setalpjdeffig}
Schematic showing the set of junctions belonging to the loop $\alpha$
(left) and the set of loops containing the junction $j$ (right).
The cubic lattice case with the main axes aligned with the Cartesian
coordinate system (middle) is chosen.
Remaining symbols are defined in the main text. 
}
\end{figure}

In this section, we revisit the method earlier presented in Ref.\
\cite{Ryc07}, dedicated to the cubic lattice, making the essential
equations weakly-dependent on the lattice topology.
In principle, the key simplifying assumption is now only the lack of dangling
edges in the lattice, i.e., we suppose that any edge $j$ (identified with the
Josephson junction) belongs to a --- at least one --- set of edges
constituting the loop $\alpha$, denoted as $\{j(\alpha)\}=\{j_1,j_2,\dots\}$,
see Fig.\ \ref{setalpjdeffig}. 
In analogy, a~set of elementary loops containing the edge $j$ is marked as
$\{\alpha(j)\}$.

The gauge-invariant phase difference between the points $1$ and $2$ is given
by
\begin{equation}
  \label{th12ginv}
  \theta_{12}=\int_{1}^{2}d{\bf l}\cdot
  \left( \nabla\phi-\frac{2\pi}{\Phi_0}{\bf A} \right),
\end{equation}
where $\nabla\phi$ is the gradient of the phase of macroscopic wave function,
${\bf A}$ is the vector potential (accounting for both static and {\em ac}
magnetic fields), and the integration takes place between the two sides of
the junction.
Taking into account the single valuedness of the wave function and the Stokes
theorem for the elementary loop ($\alpha$), one obtains the relation between
the flux piercing the loop and the phase differences on the Josephson
junctions composing the loop
\begin{equation}
  \Phi_\alpha\equiv{\bf s}_\alpha\cdot{}{\bf B}_\alpha=
  -\frac{\Phi_0}{2\pi}\sum_{j(\alpha)}\varepsilon_{j,\alpha}\theta_j. 
\end{equation}
We label the directed area of the loop as ${\bf s}_\alpha$ and the local field as
${\bf B}_\alpha$. 
Additionally, we define
\begin{equation}
\varepsilon_{j,\alpha}=
{\rm sgn}\,(\hat{\bf l}_j\cdot{\bf r}_{j,\alpha}\times{\bf s}_{\alpha}), 
\end{equation}
where ${\rm sgn}\,x$ is the sign function, 
$\hat{\bf l}_j$ is a~unit vector indicating the direction of the junction
(and the sign convention for $\theta_j$) 
and ${\bf r}_{j\alpha}={\bf r}_\alpha-{\bf r}_j$, with ${\bf r}_\alpha$
(or ${\bf r}_j$) being the position of the center of the loop (or junction). 

Introducing the self-inductance of the loop $L_\alpha$ and the loop circular
current $I_\alpha$, we can write down the flux as
\begin{equation}
\Phi_\alpha=\Phi_\alpha^{\rm ext}-L_\alpha{}I_\alpha,\ \ \ 
\text{with }\ \ \ 
\Phi_\alpha^{\rm ext}={\bf s}_\alpha\cdot{}{\bf B}_\alpha^{\rm ext}. 
\end{equation}
We further assume the uniform external field, with the perpendicular
microwave and quasistatic components, namely, 
\begin{equation}
\label{bealpext}
{\bf B}_\alpha^{\rm ext}(t)=B_\mu(t)\hat{\bf x}+\gamma{}t\hat{\bf z}, 
\end{equation}
with the unit vectors in the Cartesian coordinates $\hat{\bf x}$,
$\hat{\bf y}$, $\hat{\bf z}$, and the symbols $B_\mu(t)$, $\gamma$,
to be specified later. 
In turn, the loop current can be expressed as follows
\begin{equation}
\label{ialpphi}
I_\alpha=\frac{1}{L_\alpha}\left(
\frac{\Phi_0}{2\pi}\sum_{j(\alpha)}\varepsilon_{j,\alpha}\theta_j-\Phi_\alpha^{\rm ext} 
\right). 
\end{equation}

Next, the equation of motion for a~single junction, see Eq.\ (\ref{eqmo1jj}), 
can be rewritten taking into account that the bias current is composed of the
contributions from the loops attached to it, $\{\alpha(j)\}$, and that
the junction parameters ($C_j$, $R_j$, $I_{c,j}$) vary between the junctions, 
\begin{equation}
\label{eqmoialp}
C_j\frac{\Phi_0}{2\pi}\ddot{\theta}_j+\frac{1}{R_j}\frac{\Phi_0}{2\pi}\dot{\theta}_j+I_{c,j}\sin\theta_j=\sum_{\alpha(j)} \varepsilon_{j,\alpha}I_\alpha. 
\end{equation}
Unlike in the effective model for a~single junction (see Sec.\ \ref{batheor}),
the dependence of the junction critical current ($I_{c,j}$) on the flux
is now neglected.
This is because for large lattices and low fields the interference effects, 
occurring for supercurrents following different parallel paths in the lattice,
generally prevail over the single-junction diffraction-like effect. 

Eqs.\ (\ref{ialpphi}) and (\ref{eqmoialp}) define a~closed set of dynamic
equations for $\theta_j$. Notice that zero net current flowing through
the lattice is assumed, so the loop currents ($I_\alpha$) are uniquely defined
--- in Eq.\ (\ref{ialpphi}) --- via $\theta_j$ and external
$\Phi_\alpha^{\rm ext}$.
In a~finite lattice, the external flux is screened predominantly by the
self-inductance and Josephson supercurrents near the surface; these currents
govern internal supercurrent distribution. 

For the purpose of a~numerical simulation, it is useful to define the
dimensionless variables, 
\begin{align}
  \tilde{\theta}_j &\equiv{} \frac{\theta_j}{2\pi}, \ \ \ \
  \tilde{\Phi}_\alpha \equiv{} \frac{\Phi_\alpha}{\Phi_0}, \ \ \ \
  \tilde{I}_\alpha \equiv{} \frac{I_\alpha}{I_c^{\rm av}},
  \nonumber\\
  \tilde{L}_j &\equiv{} \frac{L^{\rm av}I_{c,j}}{\Phi_0},  \ \ \ \
  \tilde{L}_\alpha \equiv{} \frac{L_\alpha{}I_{c}^{\rm av}}{\Phi_0},  \ \ \ \
  \tilde{\Gamma}_j \equiv{} \frac{\sqrt{L^{\rm av}C^{\rm av}}}{R_jC_j},
  \label{dimlessdef} \\
  \tilde{t} &\equiv{} \frac{t}{\sqrt{L^{\rm av}C^{\rm av}}},  \ \ \ \
  \dot{\tilde{\theta}}_j \equiv \frac{d\tilde{\theta}}{d\tilde{t}}, \ \ \ \
  \ddot{\tilde{\theta}}_j \equiv \frac{d^2\tilde{\theta}}{d\tilde{t}^2},
  \nonumber
\end{align}
with the expectation values $I_c^{\rm av}$, $L^{\rm av}$, $C^{\rm av}$,
introduced to uniquely define the units of time and current for the whole
system. 
(Notice that the dimensionless inductance is defined separately for the
junctions and the loops, $\tilde{L}_j$, $\tilde{L}_\alpha$ to account for
the spatial variation of the physical $I_{c,j}$ and $L_\alpha$; from
definition, the expectation values are equal, i.e., 
$\tilde{L}_j^{\rm av}=\tilde{L}_\alpha^{\rm av}\equiv{}\tilde{L}_0$.)
Eqs.\ (\ref{ialpphi}) and (\ref{eqmoialp}) can now be rewritten as
\begin{align} 
\tilde{L}_\alpha\tilde{I}_\alpha &= 
\sum_{j(\alpha)}\varepsilon_{j,\alpha}\tilde{\theta}_j-\tilde{\Phi}_\alpha^{\rm ext},
\label{ialpphired}
\\
\ddot{\tilde{\theta}}_j+\tilde{\Gamma}_j\dot{\tilde{\theta}}_j
&+\tilde{L}_j\sin{}2\pi\tilde{\theta}_j
= \sum_{\alpha(j)} \varepsilon_{j,\alpha}\tilde{L}_\alpha\tilde{I}_\alpha.
\label{eqmoialpred} 
\end{align}

\section{Numerical results for disordered cubic lattice} 
\label{numres}

\begin{table*}[!bt]
\caption{
  The estimated values of the sample parameters and the corresponding values 
  of the parameters taken in the network simulation.}
\label{tab1}
\begin{tabular}{ll}
  \hline\hline
  $\ \ $Simulation parameters (dimensionless)$\ \ \ \ \ \ $
  & Sample parameters (physical units) \\
  \hline
  $\ \ $Lattice parameter:  $\tilde{s}_0=1$
  & Average grain size: $d=0.74\ \mu\mbox{m}$ \\  
  $\ \ $Microwave amplitude:  $A_m=0.1$
    &  $3.6\cdot 10^{-4}\mbox{ T}$ \\
  $\ \ $Applied field range: $\tilde{\Phi}^{\rm ext}_z=0\div 10$
    &  $B_z=0\div 0.036\mbox{ T}$                        \\
  $\ \ $Time step: $\Delta\tilde{t}=0.1$
    &  $1.06\cdot 10^{-13}\mbox{s}$  \\
  $\ \ $Microwave field period: $\tilde{T}=50$
    & Microwave frequency: $\nu=190\mbox{ GHz}$ \\
  $\ \ $Relative self-inductance: $\tilde{L}\equiv{LI_c}/{\Phi_0}=1.0\ \ \ $
  & Critical current: $j_c={I_c}/{d^2}=1.7\cdot 10^{10}\mbox{ A/}\mbox{m}^2$ \\
  $\ \ $Damping: $\tilde{\Gamma}=5.0$
    & Normal state resistivity:$^a$
    $\rho_n=R{d^2}/{\Delta}=2.4\cdot 10^{-4}\ \Omega\mbox{m}\ \ \ $ \\ 
  \hline\hline
  \multicolumn{2}{l}{
  $\ \ ^a\,\Delta=10\,$\AA$\ $is the approximate junction width.}
\end{tabular}
\end{table*}

\subsection{Details of the numerical procedure}
Following Ref.\ \cite{Ryc07}, we now limit our discussion to the case
of a~cubic lattice containing $N\times{}N\times{}N$ junctions. 
Eqs.\ (\ref{ialpphired}), (\ref{eqmoialpred}) define the system 
of $N_J=3N(N+1)^2$ second order differential equations for $\tilde{\theta}_j$,
whose solution is to be discussed numerically. 
Although the lattice topology is homogeneous, the random variations of the
junction (and loop) parameters are introduced to make the situation more
realistic. 

The physical situation of the EPR (or MMMA) measurement is reflected by
assuming that we have a microwave field of period $T$ and amplitude $A_m$ 
applied along the $x$-axis and $y$-axis and a~(quasi)static magnetic field
applied in the $z$-direction.
Assuming that the lattice axes are aligned with the coordinate system,
we can now rewrite Eq.\ (\ref{bealpext}) defining the external magnetic
field, in the dimensionless units introduced in Eq.\ (\ref{dimlessdef}), 
as follows
\begin{equation}
\label{tilphiext}
\tilde{\Phi}_\alpha^{\rm ext}(\tilde{t})=
\tilde{s}_\alpha\times
\begin{cases}
\tilde{A}_m\sin(2\pi{}\tilde{t}/\tilde{T}) 
  & \text{if }\ \ {\bf s}_\alpha||\,\hat{\bf x}, \\
0 & \text{if }\ \ {\bf s}_\alpha||\,\hat{\bf y}, \\
\tilde{\gamma}\tilde{t} 
  & \text{if }\ \ {\bf s}_\alpha||\,\hat{\bf z},  
\end{cases}
\end{equation}
where the scaling factor $\tilde{s}_\alpha$ (see below)
characterizes the size of elementary loop $\alpha$, and 
$\tilde{\gamma}\ll{}1$ describes the slow sweeping rate of the
quasistatic field.

The description of real systems requires taking into account the random
variation of the parameters of each loop in the array.
For this purpose we introduce, in Eq.\ (\ref{tilphiext}), the coefficients
$\{\tilde{s}_\alpha\}$ (with the mean $\tilde{s}_0=1$), which scale the
magnetic fluxes piercing elementary loops.
These coefficients, together with the junction parameters 
$\tilde{L}_j$, $\tilde{\Gamma}_j$, etc., fluctuate according 
to the Gaussian distribution with $10\%$ dispersion to approximate the
experimentally estimated dispersion of crystallite size \cite{Sta87} around 
the values $\tilde{s}_0\approx{}1$ (corresponding to the average grain size 
$d=0.74\,\mu\text{m}$), $\tilde{L}_0=1$, and $\tilde{\Gamma}_0=5$ 
(overdamped regime).
The representative values of the parameters in both physical and dimensionless
units (used in the simulation) are provided in Table~\ref{tab1}. 
In particular, the microwave field period $\tilde{T}=50$ 
corresponds to the frequency much higher than 
that in the experiments \cite{Sta87,Dur87,Bla87} ($9.4\,\text{GHz}$). 
The higher frequency was taken to accelerate the computations and we believe
this factor does not influence the output character in any essential way.

The system of $N_J$ differential 
equations, generated for $N=1\div{}30$ was
solved numerically using the fourth-order
Runge-Kutta method \cite{Bur11} taking randomly 
initial values of $\tilde{\theta}(0)$ and $\dot{\tilde{\theta}}(0)$, 
as well as setting the time step $\Delta\tilde{t}=0.1$. 
The sweeping rate of the quasistatic field was $\tilde{\gamma}=10^{-3}$,  
and in the range $0\leqslant\tilde{\Phi}_z^{\rm ext}\leqslant{}10$. 
Typically, the solution simulation consisted of $10^5$ time steps.

\begin{figure*}[!t]
\includegraphics[width=0.9\linewidth]{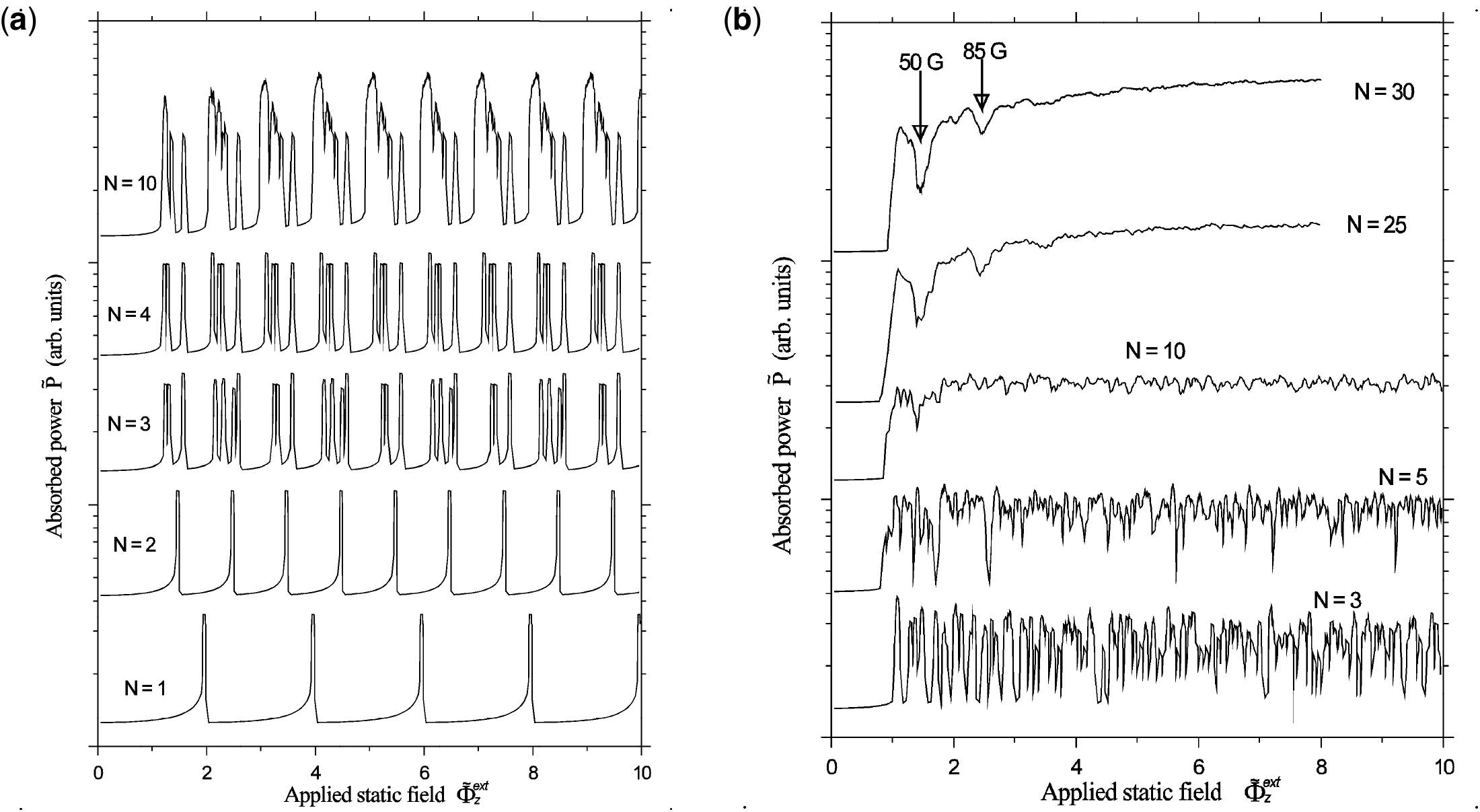}
\caption{ \label{rysivvfig}
Power absorption vs applied magnetic field for the network containing
$N\times{}N\times{}N$ Josephson junctions.
(a) Different networks of identical junctions ($\tilde{\Gamma}_0=5$,
$\tilde{L}_0=1$). 
(b) The parameters $\tilde{\Gamma}$ and  $\tilde{L}$ were undergoing $10$\%
Gaussian fluctuations around the mean values $\tilde{\Gamma}_0$,
$\tilde{L}_0$. 
(Adapted from Ref.\ \cite{Ryc07}.) 
}
\end{figure*}

\subsection{Analysis of the results}
The power absorbed (per one junction) can be written in dimensionless
units as
  \begin{equation}
  \label{pabs}
    \tilde{P}=
    \frac{1}{N_J}\left<\sum_{j=1}^{N_J}
    \tilde{\Gamma}^{x}_{j}\left(\dot{\tilde{\theta}}_{j}\right)^2
    \right>,
  \end{equation}
where averaging takes place over the time interval equal to the microwave
field period $\tilde{T}$.
This expression corresponds to the familiar expression $P=U^2/R$ in
dimensionless units, where $U$ is the voltage drop on resistance the $R$.
The representative shape of the applied field dependence (in units
$\Phi/\Phi_0$) of $\tilde{P}$ is displayed in Fig.\ \ref{rysivvfig}.

First, to understand the power-spectrum evolution with the system size, 
one has to analyze the array with variable $N$ and identical physical
parameters.
The results of the simulation for such network containing
$N\times N\times N$ junctions are displayed in Fig.\ \ref{rysivvfig}(a), 
for the values of parameters $N = 1 \div 10$, $\tilde{\Gamma}=5$
and $\tilde{L}=1$. 
With increasing $N$, we observe the systematic
filling of the space between the discrete
resonances (a~physical discussion of which is provided
below), which eventually, for $N\rightarrow\infty$, smooths out
the absorption curve above the threshold 
value of $\tilde{\Phi}^{\rm ext}_{z}=1$.
(The situation with a lack
of sizable statistical variations of the junction parameters
may have some relevance to the observed \cite{Nie97} smooth variation 
of the power absorption in YBaCuO samples.) 

Elementary analysis (see Sec.\ \ref{batheor}) suggests that 
the first absorption maximum appears for a single junction at
$\tilde{\Phi}^{\rm ext}_z=\tilde{L}$, which determines the connection 
between $\tilde{L}$ and the lower critical field for the sample
(in physical units)
  \begin{equation}
  \label{Bc1}
    B_{c1}=\frac{\tilde{L}\Phi_0}{d^2}, 
  \end{equation}
as for $\tilde{\Phi}^{\rm ext}_z=\tilde{L}$ the current screening the external
flux in the $xy$ plane achieves a critical value for
the appearance of nonzero voltage.
The numerical results for small systems presented in Fig.\ \ref{rysivvfig}(a), 
prove that the
maxima in question appear at higher field
depending. The presence of the microwave
field does not account for the difference,
as it shifts the maximum the opposite way.

The qualitative analysis of the differential equations describing the system
of $12$ identical junctions forming a~single cube ($N=1$)
leads to the following critical values of the flux 
\begin{align}
    \tilde{\Phi}_c(A_m\rightarrow 0) &=
    1+\frac{2}{\pi}\arcsin\frac{1}{\pi\tilde{L}+1} \nonumber \\
  &+ \tilde{L}\sqrt{1-\frac{1}{(\pi\tilde{L}+1)^2}}+2k,
  \label{j12fic}
\end{align}
for $k=0,1,2,\dots$ \cite{Ryc07}.
The agreement of this result with the absorption maxima obtained from
the simulation is quite good, and
improves for larger $\tilde{L}$. 
The derivation of Eq.\ (\ref{j12fic}), complemented with some numerical
examples for $\tilde{L}>1$, is given in Appendix~\ref{app12jj}. 

Fig.\ \ref{rysivvfig}(b) displays the power absorption for several
values of $N=3\div{}30$ and the junction parameters undergoing $10$\%
Gaussian fluctuations around the mean values ($\tilde{\Gamma}_0=5$,
$\tilde{L}_0=1$).
The discrete absorption peaks average out with increasing $N$; however, 
two characteristic features survive.
First, the absorption starts at the value
$\tilde{\Phi}^{\rm ext}_z=0.9$ (corresponding to $10\%$ reduction of
the elementary loop size from the standard value equal 
to unity). Second, well defined minima develop 
at specific values of $\tilde{\Phi}^{\rm ext}_z$. 
These minima are also present on the experimental curve
(cf.\ Fig.\ 1, where the derivative $dP/d\Phi_z$ was measured). 
The first minimum is positioned at 
$\tilde{\Phi}^{\rm ext}_z=1.4$, which in combination with the average
grain size $d=0.74\,\mu\text{m}$ gives the magnetic induction
$B_1=50\,\text{Gs}$ in an excellent agreement with the experiment. 
The second minimum is placed at $\tilde{\Phi}^{\rm ext}_z=2.4$, 
which leads to the induction value $B_2=85\,\text{Gs}$. 

The described effect is more pronounced withincreasing $N$ and must be
common to all inhomogeneous superconductors when the statistics improves.
Additionally, the change of the absorption spectrum obtained after introducing
the random fluctuations of the junction parameters
is rather easy to grasp quantitatively.
Each discrete absorption line at the position given by Eq.\ (\ref{j12fic}) 
is convoluted with the probability distribution.
The lines for large
$\tilde{\Phi}^{\rm ext}_z$ are smeared out proportionally in a wide region, 
so their sum will result in a practically constant value of the absorption.
The expression in Eq.\ (\ref{Bc1}) for the first critical
field is valid; that is why it was used in the evaluation of
the quantities listed in Table~\ref{tab1}. 
(This is simply due to the fact that for $\tilde{\Phi}^{\rm ext}_z<\tilde{L}$, 
absorption is zero, since the superconducting loops screen the external
field.)

\section{Possible implications for Josephson qubits} 
\label{qbitnet}

Since the pioneering work of Martinis, Devoret, and Clarke \cite{Mar20},
Josephson-junction-based quantum computing has achieved remarkable progress
and garnered significant scientific and technological interest \cite{Jia25}.
Typical operations on Josephson qubits (e.g., transmons) require the
application of both static and microwave fields \cite{Kra19}, and the
physical processes occurring in such a~situation are still not fully
understood \cite{Gua21,Che24}.

\begin{figure*}[!t]
\includegraphics[width=0.7\linewidth]{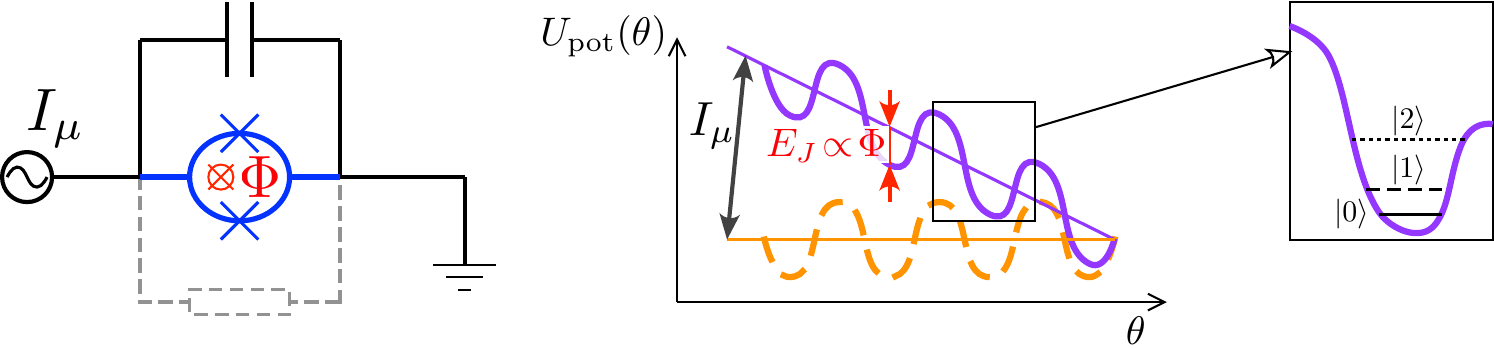}
\caption{ \label{josqubit1fig}
Left to right: 
Effective circuit for a~single Josephson qubit, the potential energy
controlled by the microwave bias current ($I_\mu$) and the magnetic flux
($\Phi$), and non-equidistant transmon energy levels ($|0\rangle$
and $|1\rangle$ span a~computational subspace). 
}
\end{figure*}

At a~first approximation, the quantum mechanics of the transmon can be
derived starting from the equation of motion for a~single junction,
see Eq.\ (\ref{eqmo1jj}), corresponding to damped oscillations governed by
the Hamiltonian 
\begin{equation}
H = \frac{Q^2}{2C}+U_{\rm pot}(\theta), 
\end{equation}
with the charge on the junction $Q=CU=C\dot{\theta}\,\Phi_0/(2\pi)$,
and the (time-dependent) potential energy
\begin{equation}
U_{\rm pot}(\theta) = I_\mu(t)\frac{\Phi_0}{2\pi}\theta-E_J\cos\theta, 
\end{equation}
where $E_J=I_c\Phi_0/(2\pi)$ is the Josephson coupling energy. 
Additionally, the critical current is controlled by replacing 
a~single junction with a~superconducting loop containing the two parallel 
junctions forming a~superconducting quantum interference device (SQUID) 
\cite{Tin04}. 
Due to the interference between the two arms enclosing the flux $\Phi$, 
see Fig.\ \ref{josqubit1fig}, 
the coupling energy is replaced by 
\begin{equation}
  E_J\rightarrow{}E_J(\Phi)=2E_J'\cos(\pi\Phi/\Phi_0), 
\end{equation}
where $E_J'$ denotes the coupling energy for a~single junction
(for simplicity, the symmetric case is considered).
For the capacity, we have $C\rightarrow{}2C'$. 

In ultra-low temperatures, the dynamical variables $Q$ and $\theta$
are replaced by quantum-mechanical operators, following the commutation
relation
\begin{equation}
\label{commquth}
  [\hat{Q},\hat{\theta}]=-2e\,i,  
\end{equation}
with the Cooper pair charge $-2e$. 
As a~result, macroscopic quantum effects appear. 
The unharmonic potential energy $U_{\rm pot}(\theta)$ results in non-equidistant
energy levels, allowing one to isolate the 
two lowest ones, $|0\rangle$ and $|1\rangle$, forming a~computational
subspace for a~transmon qubit. 
We further notice that the key finding, i.e., the commutation relation given
by Eq.\ (\ref{commquth}), can be obtained from the BSC wavefunction (see
Appendix~\ref{appbsc}). 

In other words, operations on superconducting qubits are possible since
the quantum coherence is preserved not only for the ground states, but
also for specific excited states, corresponding to charge (and phase)
oscillations for individual Josephson junctions.
In such physical regime, the role of normal-state junction resistance
is insignificant; however, nonunitary evolution, usually modeled within
the Lindblad masterequation \cite{Pel95,Din24}, occurs due to the coupling
with the enviroment. 

From this perspective, the analogy between the superconducting circuit
in present-day quantum computers and the Josephson lattice modeling
a~high-temperature superconductor may not be as far as it seems at
a~first glance. 
Very recently, Burin \cite{Bur26} pointed out that anomalous microwave
absorption in Josephson-junction qubits can be understood by referring to 
the dynamics of two-level systems in glasses. 
Since microwave absorption in Josephson networks is a~primary source of
performance-limiting decoherence for superconducting quantum computers
\cite{Moh25}, fundamental physics of these systems, in particular
a~quantum-to-classical crossover, will undoubtedly grasp a significant
attention in the next decade.

\section{Conclusions}
\label{conclu}

Selected models describing --- up to a~different degree of accuracy ---
the magnetically-modulated microwave absorption for inhomogeneous  (ceramic)
superconductors have been overviewed. 
The numerical approach, allowing one to generate the dynamic equation for
a~Josephson network with minimal assumptions on the topology, has been
presented. 

Next, {\em a~direct} numerical solution of the dynamic equation for
a~three-dimensional cubic network, with inclusion of a~microwave field, was
carried out to
describe the power absorption observed in ceramic samples of high-temperature
superconductors. 
The network of the size $30\times{}30\times{}30$ and with $10$\% random
variations of the junction parameters, including resistivity, self-inductance
and capacity, accounts well for the experimental observations. 

Finally, we put forward the analogy between Josephson networks studied
as models for high-temperature superconductors and the circuits (hosting
Josephson qubits) in superconducting quantum computers.

\section*{Acknowledgements}
I thank Professor J\'ozef Spa\l{}ek for numerous discussions.
The support by Polish supercomputing infrastructure PLGrid
(HPC Center: ACK Cyfronet AGH)  within the computational grant
No.\ PLG/2026/019777 is acknowledged.


\appendix

\section{Stability of the system of differential
          equations for the cubic elementary cell of junctions}
\label{app12jj}

In this Appendix we derive the analytic result given in Eq.\ 
(\ref{j12fic}) from the stability considerations for the system
of differential equations, describing the time evolution
of the phases. The reasoning is carried out for an 
isolated structural unit composed of $12$ 
Josephson junctions (an elementary cube), with inclusion of 
resistances and capacitances. 
This cell is aligned with the Cartesian coordinates 
and placed in the applied static field, parallel 
to the $z$-axis, and the microwave field, parallel to the $x$-axis.
Since each junction is parallel to one of the axes, we label
the phase differences by indicating the orientation of the junction
(i.e., $\theta_j^x$, $\theta_j^y$, $\theta_j^z$ with $j=1,\dots,4$);
similarly, the fluxes (currents) are labeled by specifying the relevant
plane, $xy$, $yz$, or $zx$. 

\begin{figure}[!t]
\includegraphics[width=0.6\columnwidth]{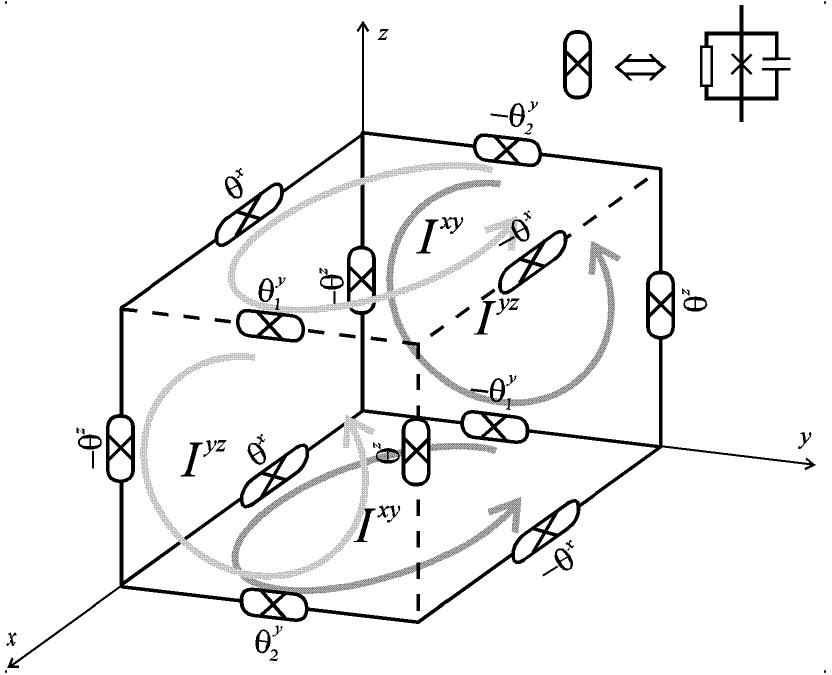}
\caption{
  Isolated cubic elementary cell composed of $12$ Josephson junctions,
  together with the corresponding independent gauge invariant 
  phases and currents. The independent magnetic fluxes 
  reflect the circular currents mark for each loop. 
}
\label{kostka}
\end{figure}

The symmetry of the problem reduces to consideration of 
four invariant phases, $\theta^x$, $\theta^z$, $\theta^y_1$, 
and $\theta^y_2$, as well as of two magnetic fluxes, $\Phi^{xy}$
and $\Phi^{yz}$, and to the loop currents $I^{xy}$ and $I^{yz}$. The
situation is depicted schematically in  Fig.\ \ref{kostka}.
The magnetic fluxes can be related to the 
phase differences and the currents in the following 
manner, see Eq. (\ref{ialpphired}), 
  \begin{equation}
  \label{kostka-fi}  
    \left.\begin{array}{c}
      \tilde{\Phi}^{xy}= - \left(\tilde{\theta}^y_2 + \tilde{\theta}^y_1 
                         + 2\tilde{\theta}^x \right) 
                       = \tilde{\Phi}^{\rm ext}_z + \tilde{L}\tilde{I}^{xy}, \\
      \tilde{\Phi}^{yz}= - \left(\tilde{\theta}^y_2 - \tilde{\theta}^y_1 
                         + 2\tilde{\theta}^z \right)  
                       = \tilde{\Phi}^{\rm ext}_x + \tilde{L}\tilde{I}^{yz}, \\
    \end{array}\right. 
  \end{equation}
where $\tilde{\Phi}^{\rm ext}_x$ and $\tilde{\Phi}^{\rm ext}_z$ are the fluxes
parallel to the corresponding axes (in units of $\Phi_0$).

The dynamic equation (\ref{eqmoialpred}) takes the form: 
  \begin{equation}
  \label{kostka-th1}
    \left.\begin{array}{c}
      \ddot{\tilde{\theta}}^x + \tilde{\Gamma}\dot{\tilde{\theta}}^x 
        + \tilde{L}\sin 2\pi\tilde{\theta}^x
      = \tilde{L}\tilde{I}^{yz} ,\\
      \ddot{\tilde{\theta}}^y_1 + \tilde{\Gamma}\dot{\tilde{\theta}}^y_1 
        + \tilde{L}\sin 2\pi\tilde{\theta}^y_1 
      = \tilde{L}\left(\tilde{I}^{xy}-\tilde{I}^{yz}\right) ,\\
      \ddot{\tilde{\theta}}^y_2 + \tilde{\Gamma}\dot{\tilde{\theta}}^y_2 
        + \tilde{L}\sin 2\pi\tilde{\theta}^y_2 
      = \tilde{L}\left(\tilde{I}^{xy}+\tilde{I}^{yz}\right) ,\\
      \ddot{\tilde{\theta}}^z + \tilde{\Gamma}\dot{\tilde{\theta}}^z 
        + \tilde{L}\sin 2\pi\tilde{\theta}^z 
      = \tilde{L}\tilde{I}^{xy} .\\
    \end{array}\right. 
  \end{equation}

Adding and substracting second and third 
equations in (\ref{kostka-th1}), and substituting expressions 
for $\tilde{L}\tilde{I}^{xy}$ and $\tilde{L}\tilde{I}^{yz}$ 
calculated from (\ref{kostka-fi}), we obtain
  \begin{equation}
  \label{kostka-th2}
    \left.\begin{array}{c}
      \ddot{\tilde{\theta}}^x + \tilde{\Gamma}\dot{\tilde{\theta}}^x 
        + \tilde{L}\sin 2\pi\tilde{\theta}^x 
      = -2\tilde{\vartheta}_+ -2\tilde{\theta}^x -\tilde{\Phi}^{\rm ext}_z, \\
      \ddot{\tilde{\vartheta}}_+ + \tilde{\Gamma}\dot{\tilde{\vartheta}}_+
        + \tilde{L}\cos 2\pi\tilde{\vartheta_-}\sin 2\pi\tilde{\vartheta}_+ 
      = -2\tilde{\vartheta}_+ -2\tilde{\theta}^x -\tilde{\Phi}^{\rm ext}_z, \\
      \ddot{\tilde{\vartheta}}_- + \tilde{\Gamma}\dot{\tilde{\vartheta}}_-
        + \tilde{L}\cos 2\pi\tilde{\vartheta_+}\sin 2\pi\tilde{\vartheta}_- 
      = -2\tilde{\vartheta}_- -2\tilde{\theta}^z -\tilde{\Phi}^{\rm ext}_x, \\
      \ddot{\tilde{\theta}}^z + \tilde{\Gamma}\dot{\tilde{\theta}}^z 
        + \tilde{L}\sin 2\pi\tilde{\theta}^z 
      = -2\tilde{\vartheta}_- -2\tilde{\theta}^z -\tilde{\Phi}^{\rm ext}_x, \\
    \end{array}\right. 
  \end{equation}
where the new variables are defined as
  \begin{equation}
  \label{def-varth}
    \tilde{\vartheta}_+\equiv
      \frac{1}{2}(\tilde{\theta}^y_2+\tilde{\theta}^y_1), 
    \ \ \ \ \ \ 
    \tilde{\vartheta}_-\equiv
      \frac{1}{2}(\tilde{\theta}^y_2-\tilde{\theta}^y_1). 
  \end{equation}

The last two equations in (\ref{kostka-th2}) describe 
damped oscillations of two coupled nonlinear oscillators
under the influence of the external force
  \begin{equation}
  \label{kostka-F}
    \underline{F}=\left(\begin{array}{c} F_{\tilde{\vartheta}_-} \\ 
                                   F_{\tilde{\theta}^z}  \\
            \end{array}\right), 
    \ \ \ \ \ \ 
    F_{\tilde{\vartheta}_-}= F_{\tilde{\theta}^z}= -\tilde{\Phi}^{\rm ext}_{x}; 
  \end{equation}
and in the potential
\begin{align}
    V\left(\tilde{\vartheta}_-, \tilde{\theta}^z\right) &=
    -\frac{\tilde{L\cos 2\pi \tilde{\vartheta}_+}}{2\pi}
    \cos 2\pi\tilde{\vartheta}_- 
\nonumber \\
    &+\left(\tilde{\vartheta}_-\right)^2
    +2\tilde{\vartheta}_-\tilde{\theta}^z +\left(\tilde{\theta}^z\right)^2
    -\frac{\tilde{L}}{2\pi}\cos 2\pi\tilde{\theta}^z.
    \label{kostka-V}
  \end{align}

We show below that the $\tilde{\vartheta}_+$ dependence
of the potential is connected to $\tilde{\Phi}^{\rm ext}_z$ and 
thus vanishes if $\tilde{\Phi}^{\rm ext}_z=0$. Therefore, 
$V\left(\tilde{\vartheta}_-, \tilde{\theta}^z\right)$ 
has a sharp minimum 
at $\tilde{\vartheta}_-=\tilde{\theta}^z=0$. Furthermore, since along
the $x$-axis $\tilde{\Phi}^{\rm ext}_x=A_m\sin\left(2\pi\tilde{t}/T\right)$,
with $A_m<<1$, and the force is given by (\ref{kostka-F}), we have for small
vibrations around minimum that
\begin{widetext}
  \begin{equation}
  \label{kostka-Vlin}
    V\left(\tilde{\vartheta}_-, \tilde{\theta}^z\right)\cong
    \frac{1}{2}\left(\begin{array}{cc}
                     \tilde{\vartheta}_- & \tilde{\theta}^z \\
               \end{array}\right)
    \left(\begin{array}{cc}
      2\pi\tilde{L}\cos 2\pi\tilde{\vartheta}_+ +2 & 2 \\
      2 & 2\pi\tilde{L}+2 \\
    \end{array}\right)
    \left(\begin{array}{c}
      \tilde{\vartheta}_- \\ \tilde{\theta}^z \\
    \end{array}\right) .
  \end{equation}
\end{widetext}
The instability threshold of the system (\ref{kostka-th2})
is reached when the quadratic form (\ref{kostka-Vlin}) 
ceases to be positively determined. This 
happens for
  \begin{equation}
  \label{kryt-varth}
    \tilde{\vartheta}_+=-\frac{1}{4}
    -\frac{1}{2\pi}\arcsin\frac{1}{\pi\tilde{L}+1}. 
  \end{equation}
It remains still to determine the relation
between $\tilde{\vartheta}_+$ and $\tilde{\Phi}^{\rm ext}_z$. 
Since $A_m\ll{}1$, then we can set at $t=0$ $\tilde{\Phi}^{\rm ext}_z=0$; then
  $
    \tilde{\theta}^y_1\cong\tilde{\theta}_y^2\cong\tilde{\theta}^x
  $, 
and hence
  \begin{equation}
  \label{przybl-varth}
    \tilde{\vartheta}_+\cong\tilde{\theta}^x, \ \ \ \ \ \ 
    \tilde{\vartheta}_-\cong 0. 
  \end{equation}
Thus, the relation to $\tilde{\Phi}^{\rm ext}_z$ can be determined 
by considering the equilibrium points, for which 
e.g. the first equation of (\ref{kostka-th2}) reduces to
the form
  $
    \tilde{L}\sin 2\pi\tilde{\theta}^x +2\tilde{\vartheta}_+ 
    +\tilde{\theta}^x= -\tilde{\Phi}^{\rm ext}_z
  $.
Taking into account (\ref{przybl-varth}) and (\ref{kryt-varth}) 
we obtain the critical flux value
  \begin{equation}
  \label{kryt-Phi}
    \tilde{\Phi}_c= 1+\frac{2}{\pi}\arcsin\frac{1}{\pi\tilde{L}+1}
    +\tilde{L}\sqrt{1-\frac{1}{(\pi\tilde{L}+1)^2}}.
  \end{equation}
This formula determines the applied field 
value, for which the first catastrophe takes place. 
Formally, the critical value of $\tilde{\vartheta}_+$, for which
the determinant of the quadratic form (\ref{kostka-Vlin}) 
vanishes, repeats every unit. In effect, the
critical value, $\Phi_c$, should repeat itself every
$4$ units. This does not happen because at 
the catastrophe, at which (\ref{kostka-V}) changes locally 
from paraboloidal into saddle-like, the 
small oscillation approximation loses its meaning.
The $\tilde{\vartheta}_-$ value does not grow without limit, 
but changes abruptly by the amount $\pm\frac{1}{2}$, 
i.e., the system jumps to the closest
minimum, where again we can use for
the time being the small-oscillation
considerations. The sign is not important. 
Nonetheless, such change of $\tilde{\vartheta}_-$ 
determines the sign change of the term
$\cos\left(2\pi\tilde{\vartheta}_-\right)$ 
in the second equation of (\ref{kostka-th2}) system. 
Therefore,  the equilibrium condition for 
$\tilde{\theta}^x$ and $\tilde{\vartheta}_+$ 
has now the form
\begin{align}
      \tilde{L}\sin 2\pi\tilde{\theta}^x +2\tilde{\vartheta}_+ 
      +2\tilde{\theta}^x &= -\tilde{\Phi}^{\rm ext}_z,
  \nonumber\\ 
      -\tilde{L}\sin 2\pi\tilde{\vartheta}_+ +2\tilde{\vartheta}_+ 
      +2\tilde{\theta}^x &= -\tilde{\Phi}^{\rm ext}_z,
\end{align}
from which we have
  $
    \sin 2\pi\tilde{\theta}^x +\sin 2\pi\tilde{\vartheta}_+ = 0
  $,
and thus, the magnitude of the jump is
  $
    \tilde{\vartheta}_+\cong\tilde{\theta}^x\rightarrow 
    \tilde{\vartheta}_+\cong\tilde{\theta}^x\pm\frac{1}{2}
  $.
This means that at each catastrophe the
system overcomes --- at a~given magnetic field --- 
half of the way to the next critical point 
(or, pulls back by the same value). This, 
in turn, leads to the frequency doubling, 
with which catastrophes happen with a slow
increase of static external field. In effect, 
this leads to
  \begin{equation}  
  \label{kryt-Phi2}
    \tilde{\Phi}_c= 1+\frac{2}{\pi}\arcsin\frac{1}{\pi\tilde{L}+1}
    +\tilde{L}\sqrt{1-\frac{1}{(\pi\tilde{L}+1)^2}}+2k,
  \end{equation}
where $k=0,1,2,...\ $.
The above formula is Eq.\ (\ref{j12fic}) from the main text. 
This approximate analytic result for $A_m\rightarrow{}0$ is compared with
the simulations for $A_m=10^{-15}$, $A_m=0.1$, and different values of
$\tilde{L}$ in Table~\ref{tab2}.

\begin{table}[!t]
\caption{
  Self-inductance dependence of the first absorption maximum position
  for an isolated cube composed of 12 Josephson junctions.  
  The values obtained from Eq.\ (\ref{j12fic})) are 
  compared with the results of computer simulation for a~very small
  ($10^{-15}$) and realistic ($0.1$) amplitudes of the microwave field. 
}
\label{tab2}
\begin{tabular}{cccc}
  \hline\hline
  $\tilde{L}$
    & $\ \,$Eq.\ (\ref{j12fic})$\ \,$
    & $\ \,A_m=10^{-15}\ \,$ & $\ \,A_m=0.1\ \,$\\
  \hline
  $\ \,$0.2$\ \,$ & 1.579 & 1.616 & 1.276 \\
  0.5 & 1.715 & 1.733 & 1.489 \\
  1.0 & 2.126 & 2.153 & 1.935 \\
  2.0 & 3.069 & 3.093 & 2.906 \\
  5.0 & 6.029 & 6.032 & 5.882 \\
  \hline\hline
\end{tabular}
\end{table}


\section{The charge-phase commutation relation}
\label{appbsc}

In this Appendix, we show how the commutation relation given by
Eq.\ (\ref{commquth}) can be easily derived by writing the BCS
wavefunction (up to the normalization) as a~coherent state of
Cooper pairs, namely 
\begin{equation}
\label{phibsc}
  |\Phi_{\rm BSC}\rangle=
  \prod_{\bf k}\left(1+\phi_{\bf k}{}c_{k\uparrow}^\dagger{}
  c_{-{\bf k}\downarrow}\right)|0\rangle=\sum_n\frac{1}{\sqrt{n!}}|n\rangle, 
\end{equation}
where
\begin{equation}
  |n\rangle=\frac{1}{\sqrt{n!}}(\Lambda^\dagger)^n|0\rangle,
  \ \ \ 
  \text{with }\ \
  \Lambda^\dagger=\sum_{\bf k}\phi_{\bf k}{}c_{{\bf k}\uparrow}^\dagger{}c_{-{\bf k}\downarrow}, 
\end{equation}
is a state containing $n$ Cooper pairs.
Theremaining symbols in Eq.\ (\ref{phibsc}) are the complex
amplitudes $\phi_{\bf k}$, with the momenta ${\bf k}$, $-{\bf k}$,
in the vicinity of the Fermi surface, and the fermionic
creation (annihilation) operator $c_{{\bf k}\sigma}^\dagger{}$ ($c_{{\bf k}\sigma}{}$),
with $\sigma=\uparrow,\downarrow$ for the two spin orientations. 

Defining the gauge transformation
$|\Phi_{\rm BSC}\rangle\rightarrow|\Phi_{\rm BSC}'(\alpha)\rangle$ via the
substitution $c_{{\bf k}\sigma}^\dagger\rightarrow e^{i\alpha}c_{{\bf k}\sigma}^\dagger$ 
immediately leads to the number operator
\begin{equation}
\hat{N}|\alpha\rangle=
\sum_n\frac{1}{\sqrt{n!}}2ne^{2in\alpha}|n\rangle=
-i\frac{d}{d\alpha}|\alpha\rangle
\end{equation}
and to $[\hat{\alpha},\hat{N}]=i$.
Since the charge of electrons in the normal state ($Q_n$) is a~good
quantum number, we can write down, for the charge-phase commutator,
$[\hat{\alpha},\hat{Q}]=[\hat{\alpha},\hat{Q}_s]$.

Introducing the two superconductors ($1,2$) forming the junction (and
the capacitor) yields the phase difference and the net charge, respectively, 
\begin{equation}
\hat{\theta}=\hat{\alpha}_1-\hat{\alpha}_2, \ \ \ \ 
\text{and }\ \ \ \ 
\hat{Q}_s = \frac{1}{2}\left(\hat{Q}_{s,1}-\hat{Q}_{s,2}\right). 
\end{equation}
Since $[\hat{\alpha}_i,\hat{Q}_{s,j}]$ for $i\neq{}j$, the commutator
\begin{equation}
[\hat{\theta},\hat{Q}_s] =
\frac{1}{2}\left( [\hat{\alpha}_1,\hat{Q}_{s,1}] 
+ [\hat{\alpha}_2,\hat{Q}_{s,2}] \right).
\end{equation}
Taking $\hat{Q}_{s,1}=-2e\hat{N}_1$ and $\hat{Q}_{s,2}=-2e\hat{N}_2$, we arrive
to Eq.\ (\ref{commquth}) in the main text.



\end{document}